\documentclass[12pt,showpacs,epsf,psfig,amsmath,amssymb,aps]{revtex4}
\usepackage{epsfig}
\usepackage{graphicx}

\def\nc{\newcommand}
\nc\CMP[1]{Commun. Math. Phys. {\bf #1}}
\nc\CP[1]{Chin. Phys. {\bf #1}}
\nc\CPL[1]{Chin. Phys. Lett. {\bf #1}}
\nc\CQG[1]{Class. Quantum Grav. {\bf #1}}
\nc\GRG[1]{Gen. Relativ. Gravit. {\bf #1}}
\nc\IJTP[1]{Int. J. Theor. Phys. {\bf #1}}
\nc\IJMPA[1]{Int. J. Mod. Phys. A {\bf #1}}
\nc\IJMPD[1]{Int. J. Mod. Phys. D {\bf #1}}
\nc\JMP[1]{J. Math. Phys. {\bf #1}}
\nc\MPLA[1]{Mod. Phys. Lett. A {\bf #1}}
\nc\NCB[1]{IL Nuovo Cimento B {\bf #1}}
\nc\NPB[1]{Nucl. Phys. B {\bf #1}}
\nc\PRD[1]{Phys. Rev. D {\bf #1}}
\nc\PRL[1]{Phys. Rev. Lett. {\bf #1}}
\nc\PLA[1]{Phys. Lett. A {\bf #1}}
\nc\PLB[1]{Phys. Lett. B {\bf #1}}
\nc\PTP[1]{Prog. Theor. Phys. {\bf #1}}
\nc\be{\begin{equation}}
\nc\ee{\end{equation}}
\nc\bea{\begin{eqnarray}}
\nc\eea{\end{eqnarray}}
\nc\sta{\sin\theta}
\nc\cta{\cos\theta}
\nc\sda{\sin^2\theta}
\nc\cda{\cos^2\theta}
\nc\nn{\nonumber}
\nc\p{\partial}
\nc\cD{\mathcal{D}}
\nc\dD{\mathcal{D}^{\dagger}}
\nc\cL{\mathcal{L}}
\nc\dL{\mathcal{L}^{\dagger}}
\nc\rda{\sqrt{\Delta_{\theta}}}
\nc\cE{\mathcal{E}}
\nc\idr{\int_{r_h+\varepsilon}^{r_h+N\varepsilon} dr}

\begin{document}
\begin{flushright}
USTC-ICTS 03-3 \\
gr-qc/0303076 \\
\end{flushright}

\title{\bf Entropy of a Kerr-de Sitter black hole due to arbitrary spin fields}
\author{Shuang-Qing WU\footnote{E-mail address: sqwu@ustc.edu.cn}}
\affiliation{\centerline{\it\footnotesize Interdisciplinary Center for Theoretical Study,
University of Science and Technology of China,} \\ \vskip -4pt
\centerline{\it\footnotesize Hefei, Anhui 230026, People's Republic of China}}
\author{Mu-Lin YAN\footnote{E-mail address: mlyan@ustc.edu.cn}}
\affiliation{\centerline{\it\footnotesize CCAST (World Lab), P. O. Box 8730, Beijing
100080, People's Republic of China} \\ \vskip -4pt
\centerline{\it\footnotesize and} \\ \vskip -4pt
\centerline{\it\footnotesize Interdisciplinary Center for Theoretical Study,
University of Science and Technology of China,} \\ \vskip -4pt
\centerline{\it\footnotesize Hefei, Anhui 230026, People's Republic of China}}


\begin{abstract}
The Newman-Penrose formalism is used to derive the Teukolsky master equations
controlling massless scalar, neutrino, electromagnetic, gravitino, and gravitational
field perturbations of the Kerr-de Sitter spacetime. Then the quantum entropy of
a non-extreme Kerr-de Sitter black hole due to arbitrary spin fields is calculated
by the improved thin-layer brick wall model. It is shown that the subleading order
contribution to the entropy is dependent on the square of the spins of particles and
that of the specific angular momentum of black holes as well as the cosmological
constant. The logarithmic correction of the spins of particles to the entropy relies
on the rotation of the black hole and the effect of the cosmological constant.
\end{abstract}

\noindent

\maketitle

\newpage
\baselineskip 22pt
\section*{I. INTRODUCTION}

Ever since Bekenstein \cite{JDB1} and Hawking \cite{SWH} discovered that black holes are
thermodynamic objects endowed with a temperature proportional to its surface gravity and
an entropy equal to one fourth of its surface area, much effort has been devoted to
investigating the statistical \cite{ZT}, quantum \cite{BKLS}, or dynamic \cite{FN} origin
of black hole entropy \cite{JDB2}. A proposal for the study of statistical origin of the
black hole entropy is the brick wall method (BWM) suggested by 't Hooft \cite{GtH}. In this
model, the black hole entropy is identified with the statistical-mechanical entropy of a
thermal gas of quantum field excitations outside the event horizon, which is composed of the
leading order correction (i.e., the standard Bekenstein-Hawking entropy) and the logarithmic
contributions to the black hole entropy. The subleading order corrections contain two parts,
in general: the logarithmic term from the integral in the optical space and the logarithmic
correction from the effective potential. This method, especially with the aid of the
Newman-Penrose formalism \cite{NP,SC}, has been successfully used in studies of the
statistical-mechanical entropy of scalar fields for some static black holes \cite{GtH,BWA}
and stationary axisymmetric black holes \cite{LK,HKP}. In the case of various static
spherically symmetric black holes, it has also been applied to evaluate the entropy of
spinor fields \cite{DF,SYG} and electromagnetic fields \cite{BE}. Recently the BWM has
been used to calculate the entropy of Dirac field for the Kerr(-Newman) black hole \cite{KN}.
However all these calculations did not consider the logarithmic contribution from the coupling
of the spin of particles with the rotation of black holes. This kind of spin-rotation coupling
effect appears \cite{WC1} in the Hawking thermal radiation spectrum of Dirac particles and
photons in some non-stationary Kerr(-Newman) black holes \cite{NKN}. Though the subleading
order correction had been included in Ref. \cite{SYG}, but only the logarithmic contribution
from the integral in the optical space had been considered there, the logarithmic term from
the effective potential including the quadratic spin terms still had been neglected in those
studies. The latter correction cannot be ignored in general, because it is of the same order
as the former in the high frequency approximations. The presence of a logarithmic divergence
in the entropy of the quantum scalar field has also been confirmed by other different approaches
\cite{SNS,MS0,GC}, such as the conical singularity approach, the heat kernel expansion method,
and the $\zeta$-function regularization technique, etc. As far as the rotating black hole case
is concerned, the logarithmic contribution both from the integral in the dragged optical space
and from the effective potential had been considered by using the BWM only for the scalar
field in recent research \cite{JYLK}.

Recently much attention has been paid to the quantum entropies of black holes due to higher spin
fields \cite{Hs,Li,JY,ALO}. Li, Shen and Gao, and Gao and Shen \cite{Hs} investigated the entropies
of arbitrary spin fields in various spherically symmetric black holes, but did not consider the
logarithmic contribution from the integral in the optical space nor the subleading order
correction from the effective potential including the quadratic spin terms. Li \cite{Li}
studied the entropy of one component of massless fields with spin $s = 1/2, 1$, and $2$ in
the Reissner-Nordstr\"{o}m black hole by the BWM and found that the logarithmic correction
to the entropy depends on the linear term of the spins of the particles. Jing and Yan
\cite{JY} calculated the entropy up to subleading terms of massless fields of spin $s > 0$ for
the Kerr black hole and showed that the contribution of the spins to the logarithmic terms
shall decrease the statistical-mechanical entropy of a Kerr black hole. L\'{o}pez-Ortega
\cite{ALO} extended this analysis to the Rarita-Schwinger field case but pointed out that
the entropy is increased by the logarithmic terms relating to the square of spins of particles.
However the coefficients of the quadratic spin terms in the expressions of entropy presented
in Refs. \cite{JY,ALO} are incompatible with each other and their validity is in doubt. Thus
it deserves to take the logarithmic terms to the black hole entropy from the effective potential
into account in details. How the spins of the quantum field changes the quantum entropy of a
rotating black hole is an interesting question and needs to be further clarified. The knowledge
of the entropy as a function of the spin of the field will be helpful to study the species
dependence problem of the BWM on a rotating black hole.

On the other hand, although the original BWM has contributed a great deal to the understanding
and calculation of the entropy of a black hole, there are some drawbacks in it such as the little
mass approximation and taking the term including $L^3$ ($L$ being the ``infrared cutoff'') as
a contribution of the vacuum surrounding the black hole, etc. The model is constructed on the
basis of thermal equilibrium at a large scale, so it cannot be applied to cases out of
equilibrium, such as spacetime with two horizons, for example, a Schwarzschild-de Sitter black
hole and Vaidya black hole \cite{LZ,LHZK}. However one can improve this original BWM by taking
only the entropy of a thin layer near the event horizon of a black hole into account, and
utilize this improved thin-layer brick wall method to resolve some thermal nonequilibrium
problems that can hardly be treated by the original BWM \cite{LZ,LHZK}.

For a Kerr-de Sitter spacetime with non-degenerate horizons, it has a cosmological horizon and
an outer black hole event horizon. As shown in Refs. \cite{MM,BM}, there will appear two cases:
(1) the general case that the temperatures of these two horizons are distinct. In this case, a
non-extreme Kerr-de Sitter spacetime is a thermal nonequilibrium system. When the cosmological
constant is very small, the two horizons will separate far away. Then each horizon, in principle,
can be treated as an isolated thermodynamical system. Although the total system consisted of
the two horizons is thermal nonequilibrium, the thin layer near the horizon can be taken as a
local thermal equilibrium system. The quantum entropy of such a black hole can be calculated
via this improved BWM which means that the entropy comes from a thin layer near the horizon.
By arguments on thermodynamics of a composite system, the total entropy of a non-extreme Kerr-de
Sitter spacetime is then  taken the sum of the contribution from each horizon; and (2) a special
case that the temperatures of the two distinct horizons are equal. The metric in this special
case is called the lukewarm solution \cite{MM,BM}. The lukewarm solution achieves thermodynamic
equilibrium, however it is unstable to changes in the mass of the black hole. In the lukewarm
case, the notion of local thermal equilibrium may not be used, and one can still work with the
original BWM to calculate the entropy of each horizon. It should be pointed out that the
expression of the entropy in the lukewarm case is just a special case of that obtained in our
general considerations. So we shall mainly consider the general case and leave the lukewarm
case for a special discussion.

In this paper, the entropies of non-extreme Kerr-de Sitter black holes due to higher spin fields
are taken up for consideration on which the effect of the cosmological constant and that of the
spins of particles are emphasized. The subject is important because the study of spacetimes
which are asymptotically de Sitter has received a great deal of attention recently. The recent
astrophysical observations \cite{PR} of type Ia supernovae which indicate a positive cosmological
constant \cite{OS} and the recent dS/conformal field theory (CFT) correspondence \cite{KdSCFT}
are two main motivations for studying the thermodynamics of black holes with a cosmological
constant. It should be noted that a realistic black hole may be in an asymptotically nonflat
space, thus it becomes important to investigate the effect of the cosmological constant on the
entropies of these kinds of black holes. In Ref. \cite{BW-AdS}, the renormalized black hole
entropy for the massive scalar field in a Schwarzschild-anti de Sitter space has been considered
recently by Winstanley via the original ``brick wall'' method. Although recent work \cite{IBW-DF}
has dealt with the entropy due to  massless Dirac fermionic and scalar fields in the Newman-Unti-
Tamburino (NUT) Kerr-Newman-de Sitter spacetime case, it puts emphasis on the improved brick wall
model and only cares for the leading correction to black hole entropy, i.e., the standard
Bekenstein-Hawking entropy.

The purpose of this paper is to deduce expressions of the entropy of non-extreme Kerr-de Sitter
black holes arising from arbitrary spin fields by using the improved brick-wall method and to
investigate effects of the spins of particles and the cosmological constant on the statistical
entropy. In this study, we carefully deal with the subleading order contribution to the entropy
not only from the integral in the dragged optical space but also the logarithmic term from the
effective potential including the quadratic spin terms, namely, the subleading corrections to
the entropy arising from the coupling of the spins of the particles with the rotation of the
black holes and the cosmological constant, regardless of a positive or negative one.

The paper is organized as follows. Within the Newman-Penrose formalism \cite{NP,SC}, we derive
in Sec. II the master equations governing massless Klein-Gordon scalar, Weyl neutrino, Maxwell
electromagnetic, Rarita-Schwinger gravitino, and linearized Einstein gravitational field
perturbations of the Kerr-de Sitter space. Section III is devoted to deducing expressions
of the statistical-mechanical entropy of the non-extreme Kerr-de Sitter black hole due to
arbitrary spin fields by using the thin-layer BWM. In Sec. IV, we give some arguments about
the difference with previously published results. The last section summarizes our discussions.
Appendix A separates the Teukolsky master equations and presents the Teukolsky-Starobinsky
identities in the Kerr-de Sitter geometry. In Appendix B, we calculate some integrals by
the thin-layer BWM.

\section*{II. PERTURBATIONS OF SPIN FIELDS IN THE KERR-DE SITTER SPACE}

The line element of the Kerr-de Sitter spacetime can be written in a coordinate system of
Boyer-Lindquist type as \cite{GWB,UK}
\be
 ds^2 = -\frac{\Delta_r}{\chi^2\Sigma}\Big(dt -a\sda d\varphi\Big)^2
 +\frac{\Delta_{\theta}\sda}{\chi^2\Sigma}\Big[a dt -(r^2+a^2)d\varphi\Big]^2
 +\Sigma\Big(\frac{dr^2}{\Delta_r} +\frac{d\theta^2}{\Delta_{\theta}}\Big) \, ,
\ee
where
\bea
 && \Delta_r = (r^2 +a^2)\Big(1 -\frac{r^2}{l^2}\Big) -2Mr \, , \quad
 \Delta_{\theta} = 1 +\frac{a^2}{l^2}\cda \, , \quad
 \chi = 1 +\frac{a^2}{l^2} \, , \nn \\
 && \Sigma = \rho\rho^* = r^2 +a^2\cda \, , \quad
 \rho = r +ia\cta \, , \quad \rho^* = r -ia\cta \, . \nn
\eea
Here $M$ is the mass of the black hole, $a$ is its angular momentum per unit mass, and
$\tilde{\Lambda} = 3/l^2$ is the cosmological constant. To fit with the recent astrophysical
observations \cite{PR} of type Ia supernovae, we require that the constant $\tilde{\Lambda}$
is positive but very small \cite{OS}. This means that the radius of the de Sitter horizon
is very large. The metric determinant, Ricci scalar, and the only nonvanishing Weyl scalar
are, $\sqrt{-g} = \Sigma\sta/\chi^2$, $\mathcal{R} = 4\tilde{\Lambda} = 12/l^2$,
$\tilde{\Psi}_2 = -M/\rho^{*3}$, respectively.

The metric has coordinate singularities at the roots of $\Delta_r = 0$. For a range of
parameters that satisfy the condition $l^2/3\gg M^2 > a^2$, there are four distinct real
roots, three of which are positive and correspond (in decreasing order) to the cosmological,
outer, and inner horizons of the black hole, respectively. The fourth root is negative and
non-physical. A significant case is the lukewarm solution \cite{MM,BM} which satisfies the
condition $M^2 = a^2\chi^2$. In this special case the horizons are generally distinct whereas
the surface gravity on both horizons is identical. In the allowed range of the physically
meaningful parameters $a$ and $M$, the admissible classes of the Kerr-de Sitter solutions
are categorized completely by Booth and Mann \cite{BM} in details. In this paper, we shall
assume that the horizons are non-degenerate and are mainly interested in the general case
that the temperature of the cosmological horizon and that of the outer black hole event
horizon are different from one another, and separate the lukewarm case for a special
discussion.

To derive a single master equation governing the perturbations of the Kerr-de Sitter spacetime,
we work it within the Newman-Penrose formalism \cite{NP,SC} by choosing the null-tetrad vectors
as
\bea
 && l^{\mu} = \frac{1}{\Delta_r}\Big[(r^2 +a^2)\chi,
 \Delta_r, 0, a\chi\Big] \, , \nn \\
 && n^{\mu} = \frac{1}{2\Sigma}\Big[(r^2 +a^2)\chi,
 -\Delta_r, 0, a\chi\Big] \, , \nn \\
 && m^{\mu} = \frac{1}{\sqrt{2\Delta_{\theta}}\rho}
 \Big(i\chi a\sta, 0, \Delta_{\theta}, \frac{i\chi}{\sta}\Big) \, ,
 \quad \overline{m}^{\mu} = (m^{\mu})^* \, ,
\eea
and obtain the nonvanishing spin coefficients as follows:
\bea
 && \tilde{\rho} = \frac{-1}{\rho^*} \, , \quad
 \mu = \frac{-\Delta_r}{2\Sigma\rho^*} \, , \quad
 \gamma = \mu +\frac{\Delta_r^{\prime}}{4\Sigma} \, , \quad
 \tau = \frac{-ia\rda\sta}{\sqrt{2}\Sigma} \, , \nn \\
 && \pi = \frac{ia\rda\sta}{\sqrt{2}\rho^{*2}} \, , \quad
 \beta = \frac{\rda}{2\sqrt{2}\rho}\Big(\cot\theta
 +\frac{\Delta_{\theta}^{\prime}}{2\Delta_{\theta}}\Big) \, ,
 \quad \alpha = \pi -\beta^* \, ,
\eea
where a prime denotes the partial differential with respect to its argument.

Assuming that the azimuthal and time dependence of the perturbed fields will be
of the form $e^{i(m\varphi-\omega t)}$, we find that the directional derivatives
are \cite{UK}
\bea
 && D = l^{\mu}\p_{\mu} = \cD_0 \, , \qquad\quad
 \Delta = n^{\mu}\p_{\mu} = \frac{-\Delta_r}{2\Sigma}\dD_0 \, , \nn \\
 && \delta = m^{\mu}\p_{\mu}
 = \frac{\rda}{\sqrt{2}\rho}\dL_0 \, , \quad
 \overline{\delta} = \overline{m}^{\mu}\p_{\mu}
 = \frac{\rda}{\sqrt{2}\rho^*}\cL_0 \, ,
\eea
where
\bea
 && \cD_n = \frac{\p}{\p r} -\frac{i\chi K_1}{\Delta_r}
 +n\frac{\Delta_r^{\prime}}{\Delta_r} \, , \quad
 \cL_n = \frac{\p}{\p\theta} -\frac{\chi K_2}{\Delta_{\theta}}
 +n\Big(\cot\theta +\frac{\Delta_{\theta}^{\prime}}{2\Delta_{\theta}}\Big) \, , \nn \\
 && \dD_n = \frac{\p}{\p r} +\frac{i\chi K_1}{\Delta_r}
 +n\frac{\Delta_r^{\prime}}{\Delta_r} \, , \quad
 \dL_n = \frac{\p}{\p\theta} +\frac{\chi K_2}{\Delta_{\theta}}
 +n\Big(\cot\theta +\frac{\Delta_{\theta}^{\prime}}{2\Delta_{\theta}}\Big) \, , \nn
\eea
and
\be
 K_1 = \omega(r^2 +a^2) -ma \, , \quad
 K_2 = a\omega\sta -\frac{m}{\sta} \,
\ee
with the relations
\be
 K_1 -K_2a\sta = \omega\Sigma \, , \quad
 K_2^{\prime} +K_2\cot\theta = 2a\omega\cta \, .
\ee

Using the Newman-Penrose formalism \cite{NP,SC} it can be shown that perturbation master
equations in the Kerr-de Sitter geometry are separable for massless spin $s = 0, 1/2, 1,
3/2$, and $2$ fields \cite{STU1}. The Teukolsky's master equations \cite{TP} controlling
the perturbations of Kerr-de Sitter black hole for massless arbitrary spin fields ($s =
1/2$, $1$, $3/2$, and $2$ for Weyl neutrino, source-free Maxwell electromagnetic,
Rarita-Schwinger gravitino, and the linearized Einstein gravitational fields,
respectively) read \cite{TS1}
\bea
 && \Big\{\big[D -(2s-1)\epsilon +\epsilon^* -2s\tilde{\rho}
 -\tilde{\rho}^*\big]\big(\Delta -2s\gamma +\mu\big) \nn \\
 &&\quad -\big[\delta -(2s-1)\beta -\alpha^* -2s\tau
 +\pi^*\big]\big(\overline{\delta} -2s\alpha +\pi\big) \nn \\
 &&\qquad\qquad\qquad\qquad -(s-1)(2s-1)\tilde{\Psi}_2\Big\}\Phi_s = 0 \, , \label{eq7}
\eea
for spin weight $s = 1/2, 1, 3/2, 2$ and
\bea
 && \Big\{\big[\Delta +(2s-1)\gamma -\gamma^* +2s\mu
 +\mu^*\big]\big(D +2s\epsilon -\tilde{\rho}\big) \nn \\
 &&\quad -\big[\overline{\delta} +(2s-1)\alpha +\beta^* +2s\pi
 -\tau^*\big]\big(\delta +2s\beta -\tau\big) \nn \\
 &&\qquad\qquad\qquad\qquad -(s-1)(2s-1)\tilde{\Psi}_2\Big\}\Phi_{-s} = 0 \, , \label{eq8}
\eea
for spin weight $s = -1/2, -1, -3/2, -2$ ($s$ is the spin of the perturbed fields).
It is clear that Eqs. (\ref{eq7}) and (\ref{eq8}) are also valid when $s = 0$, they
coincide with the massless minimally coupled scalar field equation
\be
\big(\Box +\mathcal{R}/6\big)\Phi = 0 \, ,
\ee
with $\Phi = \Phi_0 = \Phi_{-0}$.

All the above equations are separable by using the Newman-Penrose formalism, and
can be written as (ignoring the factor $e^{i(m\varphi-\omega t)}$)
\bea
&&\Big[\frac{1}{\Sigma}\big(\Delta_r\cD_1\dD_s +\rda\dL_{1-s}\rda\cL_s)
+2(2s-1)\Big(\frac{i\chi\omega}{\rho^*}-\frac{s-1}{l^2}\Big)\Big]\Phi_s = 0 \, ,
\label{eq10}\\
&&\Big[\frac{1}{\Sigma}\big(\Delta_r\dD_{1-s}\cD_0 +\rda\cL_{1-s}\rda\dL_s)
-2(2s-1)\Big(\frac{i\chi\omega}{\rho^*}+\frac{s-1}{l^2}\Big)\Big]
\big(\rho^{*2s}\Phi_{-s}\big) = 0 \, . \label{eq11}
\eea
It can be directly shown that Eqs. (\ref{eq10}) and (\ref{eq11}) are also satisfied
by the scalar Debye potentials $\phi_s = \Phi_s/\rho^{*2s}$ and $\phi_{-s} =
\rho^{*2s}\Phi_{-s}$, which obey \cite{TS1}
\bea
 && \Big\{\big[D -(2s-1)\epsilon +\epsilon^* -\tilde{\rho}^*\big]
 \big[\Delta -2s\gamma -(2s-1)\mu\big] \nn \\
 &&\quad -\big[\delta -(2s-1)\beta -\alpha^*
 +\pi^*\big]\big[\overline{\delta} -2s\alpha -(2s-1)\pi\big] \nn \\
 &&\qquad\qquad\qquad\qquad  -(s-1)(2s-1)\tilde{\Psi}_2\Big\}\phi_s = 0 \, , \\
 && \Big\{\big[\Delta +(2s-1)\gamma -\gamma^*
 +\mu^*\big]\big[D +2s\epsilon +(2s-1)\tilde{\rho}\big] \nn \\
 &&\quad -\big[\overline{\delta} +(2s-1)\alpha +\beta^*
 -\tau^*\big]\big[\delta +2s\beta +(2s-1)\tau\big] \nn \\
 &&\qquad\qquad\qquad\qquad -(s-1)(2s-1)\tilde{\Psi}_2\Big\}\phi_{-s} = 0 \, .
\eea
In Ref. \cite{STU1}, the authors showed that Eqs. (\ref{eq10}) and (\ref{eq11}) can be
further separated and transformed into Heun equations. In the case of a Kerr black hole,
they degenerate to the generalized spheroidal wave equation \cite{GSWE,WC2}, a confluent
form of Heun equation \cite{WC2}. Exact solutions to these equations, and integral equations
as well as other related applications can be found in Refs. \cite{STU1,GSWE,WC2,STU2,MS}.
To the end of this paper, we do not take these into account, but just present in Appendix
A the radial Teukolsky-Starobinsky identities \cite{TS1,TS2} where the coefficient $C_2$
corrects the previously published results \cite{KCM}. From their obvious expressions of
the above equations (\ref{eq10}) and (\ref{eq11})
\bea
&&\frac{1}{\Sigma}\Big\{\Delta_r^{-s}\frac{\p}{\p r}
\Big(\Delta_r^{1+s}\frac{\p}{\p r}\Big)
+\frac{\chi^2 K_1^2 -is\chi K_1\Delta_r^{\prime}}{\Delta_r}
+\frac{s}{2}\Delta_r^{\prime\prime} +\frac{1}{\sta}\frac{\p}{\p\theta}
\Big(\Delta_{\theta}\sta\frac{\p}{\p\theta}\Big) \nn \\
&&\qquad -\frac{1}{\Delta_{\theta}}\Big[\chi K_2
-s\Big(\frac{\Delta_{\theta}^{\prime}}{2}+\Delta_{\theta}\cot\theta\Big)\Big]^2
+4is\chi\omega\rho -\frac{4s^2+2}{l^2}\Sigma\Big\}\Phi_s = 0 \, , \label{eq14}
\eea
and
\bea
&&\frac{1}{\Sigma}\Big\{\Delta_r^s\frac{\p}{\p r}
\Big(\Delta_r^{1-s}\frac{\p}{\p r}\Big)
+\frac{\chi^2 K_1^2 +is\chi K_1\Delta_r^{\prime}}{\Delta_r}
-\frac{s}{2}\Delta_r^{\prime\prime} +\frac{1}{\sta}\frac{\p}{\p\theta}
\Big(\Delta_{\theta}\sta\frac{\p}{\p\theta}\Big) \nn \\
&&\qquad -\frac{1}{\Delta_{\theta}}\Big[\chi K_2
+s\Big(\frac{\Delta_{\theta}^{\prime}}{2}+\Delta_{\theta}\cot\theta\Big)\Big]^2
-4is\chi\omega\rho -\frac{4s^2+2}{l^2}\Sigma\Big\}\phi_{-s} = 0 \, , \label{eq15}
\eea
one can easily find that they are dual by interchanging $s\to -s$. Thus one only needs to
consider the case of positive spin state $p = s$, and obtain the results for the
negative spin state $p = -s$ by substituting $s\to -s$. Eqs. (\ref{eq14}) and
(\ref{eq15}) can be combined into the form of Teukolsky's master equation \cite{TP}
\bea
&&\Bigg\{\frac{\Delta_r}{\Sigma}\frac{\p^2}{\p r^2} +\frac{(1+s)\Delta_r^{\prime}}{\Sigma}
\frac{\p}{\p r} +\frac{\Delta_{\theta}}{\Sigma}\frac{\p^2}{\p\theta^2}
+\frac{\Delta_{\theta}^{\prime}+\Delta_{\theta}\cot\theta}{\Sigma}\frac{\p}{\p\theta}
+\frac{\omega^2\chi^2}{\Sigma}\Big[\frac{(r^2+a^2)^2}{\Delta_r}
-\frac{a^2\sda}{\Delta_{\theta}}\Big] \nn \\
&&\quad -\frac{2\omega ma\chi^2}{\Sigma}\Big(\frac{r^2+a^2}{\Delta_r}
-\frac{1}{\Delta_{\theta}}\Big) +\frac{m^2\chi^2}{\Sigma}\Big(\frac{a^2}{\Delta_r}
-\frac{1}{\Delta_{\theta}\sda}\Big) +\frac{2s\omega\chi}{\Sigma}\Big[a\sta
\Big(\frac{\Delta_{\theta}^{\prime}}{2\Delta_{\theta}} -\cot\theta\Big) \nn \\
&&\qquad -\frac{i\Delta_r^{\prime}}{2\Delta_r}\big(r^2+a^2\big) +2ir\Big]
+\frac{2sm\chi}{\Sigma}\Big[\frac{ia\Delta_r^{\prime}}{2\Delta_r} -\frac{1}{\sta}
\Big(\frac{\Delta_{\theta}^{\prime}}{2\Delta_{\theta}} +\cot\theta\Big)\Big]
+\frac{s}{2\Sigma}\Delta_r^{\prime\prime} -\frac{4s^2+2}{l^2} \nn \\
&&\qquad\quad -\frac{s^2\Delta_{\theta}}{\Sigma}
\Big(\frac{\Delta_{\theta}^{\prime}}{2\Delta_{\theta}}
+\cot\theta\Big)^2\Bigg\}\Psi_s = 0 \, , \label{eq16}
\qquad\qquad (s = 0, \pm 1/2, \pm 1, \pm 3/2, \pm 2).
\eea
In the above master equation (\ref{eq16}), one can identify the term proportional to
$(4s^2+2)/l^2 = (2s^2+1)\mathcal{R}/6$ with the conformally coupled term for arbitrary
spin fields in virtue of the nonvanishing Ricci scalar in the Kerr-de Sitter spacetime.
In the next section we will utilize this equation to obtain the density of states by
using the Wentzel-Kramers-Brillouin (WKB) approximation scheme.

\section*{III. ENTROPY OF KERR-DE SITTER BLACK HOLES DUE TO SPIN FIELDS}

Now we calculate the entropy due to arbitrary spin fields for the non-extreme Kerr-de Sitter
black hole by the thin-layer BWM. First we try to seek the total number of modes with energy
less than $\omega$. In order to do this, we make use of the WKB approximation and substitute
$\Psi_s \sim e^{iG(r,\theta)}$ into the above Teukolsky's master equation (\ref{eq16}), then
we obtain
\bea
&&\frac{\Delta_r}{\Sigma}k_r^2 +\frac{\Delta_{\theta}}{\Sigma}k_\theta^2
+\frac{\omega^2\chi^2}{\Sigma}\Big[\frac{a^2\sda}{\Delta_{\theta}}
-\frac{(r^2+a^2)^2}{\Delta_r}\Big] +\frac{2\omega ma\chi^2}{\Sigma}
\Big(\frac{r^2+a^2}{\Delta_r} -\frac{1}{\Delta_{\theta}} \Big) \nn \\
&&\quad +\frac{m^2\chi^2}{\Sigma}\Big(\frac{1}{\Delta_{\theta}\sda}
-\frac{a^2}{\Delta_r}\Big) +\frac{2s\omega\chi a\sta}{\Sigma}
\Big(\cot\theta -\frac{\Delta_{\theta}^{\prime}}{2\Delta_{\theta}}\Big)
+\frac{2sm\chi}{\Sigma\sta}\Big(\frac{\Delta_{\theta}^{\prime}}{2\Delta_{\theta}}
+\cot\theta\Big) \nn \\
&&\qquad\qquad +\frac{s^2\Delta_{\theta}}{\Sigma}\Big(\frac{
\Delta_{\theta}^{\prime}}{2\Delta_{\theta}} +\cot\theta\Big)^2
+\frac{4s^2+2}{l^2} -\frac{s}{2\Sigma}\Delta_r^{\prime\prime} = 0 \, ,
\label{eq17}
\eea
in which, $k_r = G_{,r}$ and $k_{\theta} = G_{,\theta}$ are the ``wave numbers.''
In terms of the covariant metric components $g_{\mu\nu}$, Eq. (\ref{eq17}) can be
rewritten as
\be
\frac{k_r^2}{g_{rr}} +\frac{k_\theta^2}{g_{\theta\theta}}
+\frac{g_{\varphi\varphi}\omega^2 +2g_{t\varphi}m\omega +g_{tt}m^2}{\cD}
+2\big(\omega B +mC\big) +H_s = 0 \, , \label{eq18}
\ee
where
\bea
&&g_{rr} = \frac{\Sigma}{\Delta_r} \, , \quad
g_{\theta\theta} = \frac{\Sigma}{\Delta_{\theta}} \, , \quad
g_{tt} = \frac{\Delta_{\theta}a^2\sda -\Delta_r}{\chi^2\Sigma} \, , \quad
g_{t\varphi} = \frac{\Delta_r -(r^2+a^2)\Delta_{\theta}}{\chi^2\Sigma}a\sda
\, , \quad \nn \\
&& g_{\varphi\varphi} = \frac{(r^2+a^2)^2\Delta_{\theta}
-\Delta_ra^2\sda}{\chi^2\Sigma}\sda \, , \quad
\cD = g_{tt}g_{\varphi\varphi}-g_{t\varphi}^2 =
-\frac{\Delta_r\Delta_{\theta}\sda}{\chi^4} \, , \nn \\
&&\qquad\qquad B = \frac{s\chi a\sta}{\Sigma}\Big(\cot\theta
-\frac{\Delta_{\theta}^{\prime}}{2\Delta_{\theta}}\Big) \, , \qquad
C = \frac{s\chi}{\Sigma\sta}\Big(\frac{\Delta_{\theta}^{\prime}}{2\Delta_{\theta}}
+\cot\theta\Big) \, , \nn \\
&& \qquad\qquad H_s = \frac{s^2\Delta_{\theta}}{\Sigma}\Big(\frac{
\Delta_{\theta}^{\prime}}{2\Delta_{\theta}} +\cot\theta\Big)^2
+\frac{4s^2+2}{l^2} -\frac{s}{2\Sigma}\Delta_r^{\prime\prime} \, .
\eea

To obtain the density of states, let us suppose that the quantum field is rotating with
an angular velocity $\Omega_h$ in a thin-layer very near the horizon of the non-extreme
Kerr-de Sitter black hole. Substituting $\cE = \omega -m\Omega_h$ into Eq. (\ref{eq18}),
we reduce it into the form
\bea
&&\frac{k_r^2}{g_{rr}} +\frac{k_\theta^2}{g_{\theta\theta}}
+\frac{g_{\varphi\varphi}\cE^2 +2(g_{t\varphi} +g_{\varphi\varphi}\Omega_h)m\cE
+\widetilde{g}_{tt}m^2}{\cD} \nn \\
&&\qquad\qquad\qquad +2\Big[\cE B +m(\Omega_hB +C)\Big] +H_s = 0 \, ,
\eea
and then rewrite it as
\be
\frac{k_r^2}{g_{rr}} +\frac{k_\theta^2}{g_{\theta\theta}}
+\frac{-\widetilde{g}_{tt}}{-\cD}\Big(m +m_0\Big)^2
= \frac{1}{-\widetilde{g}_{tt}}
\Big(\cE +s W\Big)^2 -V_s \, , \label{eq21}
\ee
where
\bea
&& \widetilde{g}_{tt} = g_{tt}+2g_{t\varphi}\Omega_h
+g_{\varphi\varphi}\Omega_h^2 \, , \nn \\
&&m_0 = \frac{(g_{t\varphi} +g_{\varphi\varphi}\Omega_h)\cE
+(\Omega_hB +C)\cD}{g_{tt}+2g_{t\varphi}\Omega_h
+g_{\varphi\varphi}\Omega_h^2} \, , \nn \\
&&s W = (g_{tt} +g_{t\varphi}\Omega_h)B -(g_{t\varphi}
+g_{\varphi\varphi}\Omega_h)C \, , \nn \\
&&V_s = H_s -(g_{tt}B^2-2g_{t\varphi}BC+g_{\varphi\varphi}C^2)
= P -\frac{s\Delta_r^{\prime\prime}}{2\Sigma} \, .
\eea
$\widetilde{g}_{tt}$ is the temporal component of the metric of the dragged
optical space \cite{AO,WC3}, $W$ is the angular velocity caused by the rotation
of the black hole and can be called the ``spin potential,'' while $V_s$ is the
effective potential \cite{LK} which is equal to $\mu^2 +\xi\mathcal{R} = \mu^2
+12\xi/l^2$ in the case of a massive scalar field with an arbitrary conformally
coupling ($\xi = 1/6$ for the minimally coupling), where $\mu$ is the mass of
the field.

For a given energy $\omega = \cE+m\Omega_h$ and a given azimuthal angular momentum
$m$, Eq. (\ref{eq21}) represents an ellipsoid of three-dimensional momentum space,
{\sl a compact surface}, spanned by $k_r$, $k_{\theta}$, and $m$, which is a subspace
of a six-dimensional phase space, supposed that the following conditions could be
satisfied:
\be
g_{rr} > 0 \, , \quad g_{\theta\theta} >0 \, , \quad
-\widetilde{g}_{tt} > 0 \, , \quad -\cD > 0 , \qquad
(\cE +s W)^2 +\widetilde{g}_{tt}V_s \geq 0 \, .
\ee
Therefore in this case, for the positive spin state of spin fields the number of
modes with $\cE$ is equal to the number of states in this classical phase space
\cite{LK}
\bea
&&\Gamma(\cE,s) = \frac{1}{(2\pi)^3} \int dr d\theta d\varphi
\int dk_r dk_{\theta}d m \nn \\
&&\qquad\quad= \frac{1}{3\pi} \int d\theta \idr
\frac{\sqrt{-g}}{(-\widetilde{g}_{tt})^2}\Big[(\cE +s W)^2
+\widetilde{g}_{tt}V_s\Big]^{3/2} \, .
\eea

Here we impose the improved thin-layer BWM boundary conditions in the integral with
respect to the radial coordinate $r$, that is, we assume that the quantum field is
equal to zero for $r \leq r_h + \varepsilon$ and $r \geq r_h + N\varepsilon$, with $r_h
\gg \varepsilon$. The ultraviolet cutoff $\varepsilon$ is a small distance from the
horizon $r_h$ to the inner brick wall, the cutoff parameter $N$ is a sufficient big
integer to remove the infrared divergence. In other words, the infrared cutoff $L$
in the original BWM is now replaced by $r_h + N\varepsilon$ in the upper limit of
the radial integral. This reflects a significant difference from the original BWM,
and such improved BWM is called the thin-layer model \cite{LHZK}, it can overcome
some defects in the original BWM.

In the original brick-wall model, it is supposed that the black hole is in thermal
equilibrium with the external field in a large spatial region. This method cannot be
applied to a nonequilibrium system such as a system of spacetime with two horizons, for
example, Schwarzschild-de Sitter spacetime and Vaidya spacetime \cite{LZ,LHZK}. In such
spacetimes, two problems arise: (a) the thermal equilibrium between the external field
and the hole is unstable, so the thermal equilibrium on a large scale basis for the brick
wall does not exist; and (b) since the two horizons have different temperatures, there
exists no global thermal equilibrium over the entire spacetime, and statistical physics
laws are invalid there.

In the thin-layer method, the entropy of the black hole is mainly attributed to the degrees of
the freedom of the field in the thin-layer ($r_h + \varepsilon \leq r \leq r_h + N\varepsilon$)
covering the surface of the horizon. The Bekenstein-Hawking entropy should be associated with
the fields in this small region near the horizon, where the local thermal equilibrium exists
and statistical physics laws are still valid. To guarantee that the notion of local thermal
equilibrium can work very well, here one supposes that the physical quantities of the
thermodynamic properties of the exciting field outside the hole vary slightly in the vicinity
of the horizon. On the one hand, the length of the thin-layer region near the horizon must
be small enough on a macroscopic scale so that the physical quantities in the region can
approximately be treated as some constants in the vicinity, and approximate equilibrium in
the small region is achieved. On the other hand, the region must be large enough on a
microscopic scale so that the statistical mechanics are valid for the fields in the near
horizon region, and the thermodynamic variables can be defined through a partition function.
In order for local equilibrium to be maintained, it is necessary that the hole's radiation
is slight enough that the fluctuation of thermodynamic properties of fields can be treated
as a small quantity. In most cases, this condition can be satisfied except for those ones
at the Planckian scale.

In the spacetime that has two horizons with two different temperatures, there exists no global
thermal equilibrium in the entire spacetime, however approximate thermal equilibrium exists
in the two layers near each horizon. Thus the global thermal equilibrium is not needed, the
validity of local thermal equilibrium is crucial to the discussion \cite{LHZK}. In such a
thin-layer BWM, the total entropy is mainly attributed to the two thin-layers near the two
horizons, namely, it is a linear sum of the area of each horizon.

In the case of a non-extreme Kerr-de Sitter black hole with non-degenerate horizons,
situation is more involved. The issue of local versus global thermal equilibrium is a rather
delicate one. As mentioned before, there are two cases that can happen: (1) In the general
case, the temperatures of the cosmological and black hole event horizons are distinct; and
(2) in the lukewarm case, both temperatures are equal. In the general, non-extreme Kerr-de
Sitter background, there is no time-like Killing vector which is well-defined in the whole
region surrounded by the two horizons. Hence the system in the general case cannot be in
thermal equilibrium globally, namely, there exists no thermal equilibrium in the entire
spacetime since the two horizons have different temperatures. In addition, there cannot
exist a global thermal equilibrium between the external field and the hole in a large spatial
region. Thus in this general case, one must work with the thin-layer method to calculate the
entropy of the Kerr-de Sitter black hole. In the lukewarm case ($M^2 = a^2\chi^2$), however,
the cosmological and black hole event horizons are in thermal equilibrium (but unstable to
changes in the mass of the black hole \cite{MM}). In this special case, one may not introduce
the notion of local thermal equilibrium and still adopt the original BWM to calculate the
entropy of each horizon. The main interest of this paper is in the general case, and we will
turn to the lukewarm case for a special consideration. In both cases, the total entropy of
a quantum field in the non-extreme Kerr-de Sitter spacetime is summed up from the entropies
corresponding to the maximal and minimal spin-weight components.

Summing over the positive and negative spin states $p= \pm s$, we get the total states number
\bea
&&\Gamma(\cE) = \frac{g_s}{2}\Big[\Gamma(\cE,s) +\Gamma(\cE,-s)\Big] \nn \\
&&\qquad = \frac{g_s}{6\pi} \int d\theta \idr
\frac{\sqrt{-g}}{(-\widetilde{g}_{tt})^2}\Bigg\{\Big[(\cE +s W)^2
+\widetilde{g}_{tt}V_s\Big]^{3/2}+ \Big[(\cE -s W)^2
+\widetilde{g}_{tt}V_{-s}\Big]^{3/2}\Bigg\} \nn \\
&&\qquad \approx \frac{g_s}{3\pi} \int d\theta \idr
\frac{\sqrt{-g}}{(-\widetilde{g}_{tt})^2}\Big[\cE^3
+3\Big(\frac{1}{2}\widetilde{g}_{tt}P+s^2W^2\Big)\cE
-\frac{3s^2\Delta_r^{\prime\prime}}{4\Sigma}\widetilde{g}_{tt}W\Big] \nn \\
&&\qquad \equiv \frac{g_s}{3\pi}\Big[I_1\cE^3 +3\big(I_2
+s^2I_3\big)\cE -3s^2I_4\Big] \, . \label{eq25}
\eea
In the above, we have expanded Eq. (\ref{eq25}) in the high frequency approximation
and introduced an appropriate degeneracy $g_s$ for each species of particles (It is
well-known that $g_s = 2s +1$ in the non-relativistic quantum statistics case). Here
$g_s = 1$ for scalar field ($s = 0$), $g_s = 2$ for Weyl neutrino ($s = 1/2$), Maxwell
electromagnetic ($s = 1$), Rarita-Schwinger gravitino ($s = 3/2$) and linearized Einstein
gravitational ($s = 2$) fields, and $g_s = 4$ for massless Dirac field ($s = 1/2$),
respectively. The following four integrals in terms of the thin-layer BWM are calculated
in Appendix B
\bea
&&I_1 = \int d\theta \idr \frac{\sqrt{-g}}{\widetilde{g}_{tt}^2} \, , \qquad
I_2 = \frac{1}{2}\int d\theta \idr \frac{\sqrt{-g}}{\widetilde{g}_{tt}}P \, , \nn \\
&&I_3 = \int d\theta \idr \frac{\sqrt{-g}}{\widetilde{g}_{tt}^2}W^2 \, , \qquad
I_4 = \int d\theta \idr \frac{\sqrt{-g}}{\widetilde{g}_{tt}}
\frac{\Delta_r^{\prime\prime}}{4\Sigma}W \, , \nn \\
&&\quad P = \frac{4s^2+2}{l^2} +\frac{s^2\Delta_{\theta}}{\Sigma}\Big(\frac{
\Delta_{\theta}^{\prime}}{2\Delta_{\theta}} +\cot\theta\Big)^2
-(g_{tt}B^2-2g_{t\varphi}BC+g_{\varphi\varphi}C^2) \, . \label{eq26}
\eea
Apparently the integral $I_1$ is related to the volume of the dragged optical space
\cite{AO,WC3}.

It is known that a ``physical space'' must be dragged by the gravitational field with an
azimuthal angular velocity $\Omega_h$ in the stationary rotating spacetime. Since we have
supposed that a quantum field in a thin-layer very near the horizon is in local thermal
equilibrium with the non-extreme Kerr-de Sitter black hole at the asymptotic temperature
$1/\beta$ measured by an observer located at the spatial infinity, it is appropriate to
assume that the quantum field is rotating with angular velocity $\Omega_h$ in this thin-layer
also. Note that here we use the asymptotic quantities rather than those measured by a
local (corotating) observer in the thin-layer. In fact the equivalence principle implies
that a thin-layer system in local thermal equilibrium has a local Tolman inverse temperature
given by $\beta_{local} = \beta\sqrt{-\widetilde{g}_{tt}}$, with $\beta$ being the asymptotic
inverse temperature \cite{LLT}. The temperature measured by this local observer is a local
temperature $\beta_{local}$, correspondingly he also measures a local (blue-shifted) energy
$\omega_{local} = \omega/\sqrt{-\widetilde{g}_{tt}}$, where the energy $\omega$ measured by
the asymptotic observer is associated with the coordinate time $t$. The quantity $\beta\omega
= \beta_{local} \omega_{local}$ is an invariant due to the first thermodynamic law, and so is
$\beta(\omega -m\Omega_h)$, etc. As the equivalence between using local quantities and using
asymptotic ones has already been proven in Ref. \cite{MI}, one can directly adopt the asymptotic
quantities, needless to make further conversions. For such a local (quasi-)equilibrium
ensemble of states of spin fields, the free energy can be expressed as follows:
\bea
&&F = \frac{(-1)^{-2s}}{\beta}\int d m \int d\omega g(\omega, m) \ln
\left[1 -(-1)^{2s}e^{-\beta(\omega -m\Omega_h)}\right] \nn \\
&&\quad = -\int d m \int d\omega \frac{\Gamma(\omega, m)}{e^{\beta(\omega
-m\Omega_h)}-(-1)^{2s}}
= -\int_0^{\infty} d\cE\frac{\Gamma(\cE)}{e^{\beta\cE} -(-1)^{2s}} \, ,
\label{eq28}
\eea
where we have made a variable substitution $\cE = \omega -m\Omega_h$. The formulas for
the density of states $g(\omega,m) = d\Gamma(\omega,m)/d\omega$ and $\Gamma(\cE) = \int
\Gamma(\cE+m\Omega_h, m) dm$ had also been used. Inserting Eq. (\ref{eq25}) into Eq.
(\ref{eq28}) and then carrying out the integral over $\cE$, we get the expression for
the free energy
\bea
&&F = -\frac{g_s}{6\pi} \int d\theta \idr \frac{\sqrt{-g}}{(-\widetilde{g}_{tt})^2}
\int_0^{\infty} \frac{d\cE}{e^{\beta\cE} -(-1)^{2s}} \nn \\
&&\qquad\quad \times \Bigg\{\Big[(\cE +s W)^2
+\widetilde{g}_{tt}V_s\Big]^{3/2}+ \Big[(\cE -s W)^2
+\widetilde{g}_{tt}V_{-s}\Big]^{3/2}\Bigg\} \nn \\
&&\quad \approx
-\frac{g_s}{3\pi}\int_0^{\infty} \frac{d\cE}{e^{\beta\cE} -(-1)^{2s}}
\Big[I_1\cE^3+3\big(I_2+s^2I_3\big)\cE-3s^2I_4\Big] \nn \\
&&\quad = -g_s\Big[2\zeta(4)\frac{15+(-1)^{2s}}{16\pi\beta^4}I_1
+\zeta(2)\frac{3+(-1)^{2s}}{4\pi\beta^2}\big(I_2 +s^2I_3\big) \nn \\
&&\qquad\qquad -\zeta(1)\frac{1+(-1)^{2s}}{2\pi\beta}s^2I_4\Big] \, ,
\eea
where $\zeta(n) = \sum\limits_{k=1}^{\infty}1/k^n$ is the Riemann zeta function,
$\zeta(4) = \pi^4/90$, $\zeta(2) = \pi^2/6$, etc.

We are now ready to obtain the entropy of the non-extreme Kerr-de Sitter black hole due
to arbitrary spin fields from the standard formula $S = \beta^2(\p F/\p\beta)$,
\bea
&&S = \frac{g_s}{2\pi}\Big[\zeta(4)\frac{15+(-1)^{2s}}{\beta^3}I_1
+\zeta(2)\frac{3+(-1)^{2s}}{\beta}\big(I_2+s^2I_3\big) \nn \\
&&\qquad\qquad -\zeta(1)\Big(1+(-1)^{2s}\Big)s^2I_4 \Big] \, .
\eea

Next, we are in a position to consider the four integrals $I_1 \sim I_4$. It is easy to
obtain the Hawking temperature and the area corresponding to the horizon $r_h$ of the
non-extreme Kerr-de Sitter black hole
\be
\beta_h^{-1} = \frac{\kappa_h}{2\pi} =
\frac{\Delta_{r_h}^{\prime}}{4\pi\chi(r_h^2 +a^2)}
= \frac{\Delta_{r_h}^{\prime}}{\chi^2A_h} \, , \qquad
A_h = 4\pi(r_h^2 +a^2)/\chi \, .
\ee
By means of  the thin-layer BWM, we take the angular velocity of a quantum field near the
horizon of the non-extreme Kerr-de Sitter black hole as $\Omega_h = a/(r_h^2 +a^2)$, and
find that only the integrals $I_1$ and $I_2$ contribute to the leading and subleading terms
in the entropy, while the integrals $I_3$ and $I_4$ can be ignored as usual because the
integral $I_3$ can be attributed to the contribution of the vacuum surrounding the hole
due to $I_3 \sim \mathcal{O}(\epsilon)$ and $I_4$ vanishes at least up to the order
$\mathcal{O}(\epsilon)$. The final expressions for these integrals are presented in
Appendix B where the ultraviolet cutoff $\varepsilon$ is replaced by the proper distance
$\eta$ from the horizon to the inner brick wall $\eta = \int_{r_h}^{r_h+\varepsilon}
\sqrt{g_{rr}}dr \approx 2(\varepsilon \Sigma_h/\Delta_{r_h}^{\prime})^{1/2}$. In order
to be comparable with the results already appeared in the literature and to simplify the
expression, we have set the new ultraviolet cutoff $\epsilon$ and infrared cutoff
$\Lambda$ by $\eta^2 = 2\epsilon^2/15$ and $N = \Lambda^2/\epsilon^2$ as did in Refs.
\cite{MS0,JY,ALO}. With utilization of the results given by Eq. (\ref{eB7}), we obtain
the statistical-mechanical entropy
\bea &&S/g_s = \pi^3\frac{15+(-1)^{2s}}{180\beta^3}I_1
+\pi\frac{3+(-1)^{2s}}{12\beta}I_2 \nn \\
&&\qquad= \frac{15+(-1)^{2s}}{90\chi(\beta\kappa_h/\pi)^3}
\Big[\frac{15(r_h^2+a^2)}{4\epsilon^2}
+\Big(1-\frac{3r_h^2+a^2}{2l^2}\Big) \ln\frac{\Lambda}{\epsilon}\Big] \nn \\
&&\qquad\quad~~+\frac{3+(-1)^{2s}}{24\chi(\beta\kappa_h/\pi)}
\Bigg\{-\frac{4(r_h^2+a^2)}{l^2}
+s^2\Big[\frac{a^2-r_h^2}{r_h^2}+\frac{r_h^2-a^2}{l^2} \nn \\
&&\qquad\qquad +\Big(\frac{r_h^2+a^2}{r_h^2}-\frac{9r_h^2+a^2}{l^2}\Big)
\frac{r_h^2+a^2}{ar_h}\arctan\big(\frac{a}{r_h}\big)\Big]\Bigg\}
\ln\frac{\Lambda}{\epsilon} \, . \label{eq31}
\eea

Assuming that the field is in the Hartle-Hawking vacuum state and taking $\beta = \beta_h$,
we get that the entropy is given by
\bea
&&S/g_s = \frac{15+(-1)^{2s}}{16}\Big[\frac{A_h}{48\pi\epsilon^2}
+\frac{1}{45\chi}\Big(1-\frac{3r_h^2+a^2}{2l^2}\Big)
\ln\frac{\Lambda}{\epsilon}\Big] \nn \\ &&\qquad\qquad
+\frac{3+(-1)^{2s}}{4}\Bigg\{-\frac{A_h}{12\pi l^2}
+\frac{s^2}{12\chi}\Big[\frac{a^2-r_h^2}{r_h^2}+\frac{r_h^2-a^2}{l^2}
\nn \\ &&\qquad\qquad
+\Big(\frac{r_h^2+a^2}{r_h^2}-\frac{9r_h^2+a^2}{l^2}\Big)
\frac{r_h^2+a^2}{ar_h}\arctan\big(\frac{a}{r_h}\big)\Big]\Bigg\}
\ln\frac{\Lambda}{\epsilon} \, . \label{eq32}
\eea
Equation (\ref{eq31}) or (\ref{eq32}) shows that the entropy of a non-extreme Kerr-de Sitter black
hole due to arbitrary spin fields consists of two parts, the leading order contribution and
the subleading order corrections, or equivalently the contribution from the integral $I_1$
in the dragged optical space and the logarithmic term $I_2$ from the effective potential
including the quadratic spin terms. The logarithmic corrections consist of the one from the
integral $I_1$ and that from the integral $I_2$, both of them are of the same order, therefore
the latter cannot be thrown away any more. It should be noted that the coefficients of the
logarithmic divergence is universal (invariant under a change in the value of the cutoff,
or even under a change in the regulator scheme). The expression of Eq. (\ref{eq32}) may
settle down the species dependence problem of the brick wall entropy on a rotating black
hole, it covers many previously obtained results.

To see what role the conformally coupling plays in the calculation, we also present the
expression of the entropy for a massive scalar field in the non-extreme Kerr-Newman-de
Sitter black hole
\bea
&& S = \frac{A_h}{48\pi\epsilon^2} +\frac{1}{45\chi}\Big[ 1-\frac{3r_h^2+a^2}{2l^2}
-\frac{3Q^2}{4r_h^2}\Big(1+\frac{r_h^2+a^2}{ar_h}\arctan\big(\frac{a}{r_h}\big)\Big)\Big]
\ln\frac{\Lambda}{\epsilon} \nn \\
&&\qquad -\frac{A_h}{24\pi}\Big(\mu^2+\frac{12\xi}{l^2}\Big)
\ln\frac{\Lambda}{\epsilon} \, ,
\label{eq33}
\eea
with an arbitrary conformally coupling constant $\xi$. It is not difficult to find that the
conformally coupling will contribute a logarithmic correction to the entropy. By setting
$N = 1$ in Eqs. (\ref{eq32}) and (\ref{eq33}), we recover the standard Bekenstein-Hawking
entropy.

\section*{IV. DISCUSSIONS}

The final result that we obtain for the entropies of arbitrary spin fields in the non-extreme
Kerr-de Sitter space deserves some remarks.

\textbf{(a)} The entropies given above have summed up the contribution from the maximal and
minimal spin-weight states of a quantum field. Under the condition that satisfies $l^2/3\gg M^2
> a^2$, the cosmological horizon separates far away from the outer black hole horizon, the
total entropy of the non-extreme Kerr-de Sitter black hole can take a linear sum of the area
of the two horizons.

The calculations here are valid both for the outer black hole event horizon case and for the
cosmological horizon case because the two horizons are in an equal position. We think it
is also valid for the black hole event horizon of the Kerr-anti de Sitter space by changing
the sign of the cosmological constant $\tilde{\Lambda}$.

\textbf{(b)} The entropies depend not only on the spins of the particles but also on the
cosmological constant except different spin fields obey different statistics. They rely
on the quadratic terms of $s^2$ and $-1/l^2$ as well as $a^2$.

\textbf{(c)} Both the contribution of the spins and that of the cosmological constant to the
entropies are in subleading order. The spins have a tendency to increase the entropies, but
the effect of a positive cosmological constant tends to decrease them.

\textbf{(d)} The logarithmic term from the spins of the particles not only depends on the
spin-rotation coupling effect but also on the coupling between the spins of particles and
the cosmological constant. Figure 1 shows that how the coefficient of the square term of the
spins for the logarithmic correction to the entropies is affected by the specific angular
momentum of the hole and the cosmological constant.

\begin{figure}[htbp]
\begin{center}
 \epsfig{figure=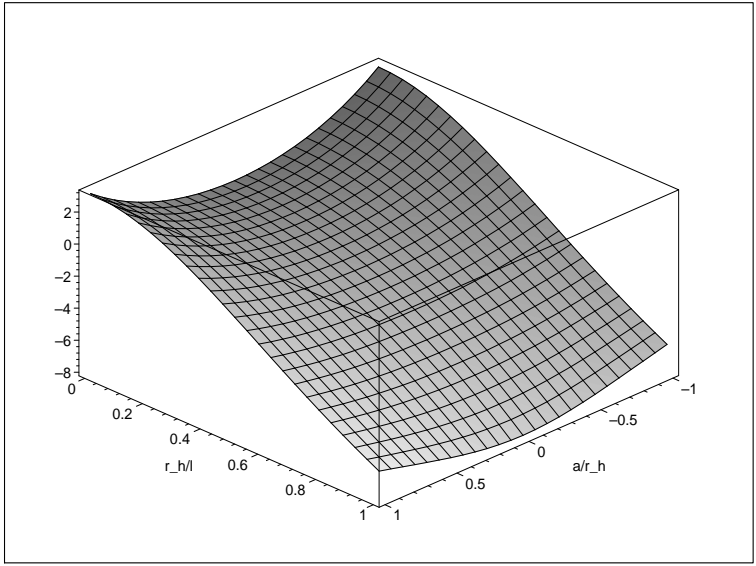,width=7.5cm,height=10cm,angle=-90}
 \caption{The coefficient of the quadratic spin term for the logarithmic
 correction to the entropies is dependent on the specific angular momentum
 of the hole and the cosmological constant. The entropy increases when
 $|a/r_h|$ becomes large and cuts down when $r_h/l$ increases.}
\end{center}
\end{figure}

\textbf{(e)} Three special cases may be very interesting:

\textbf{Case I}: In the Schwarzschild-de Sitter black hole case, the specific angular
momentum vanishes and the black hole is non-rotating. The improved thin-layer model
still works in this case, then the entropy due to arbitrary spin fields is
\bea
&&S/g_s = \frac{15+(-1)^{2s}}{16}\Big[\frac{A_h}{48\pi\epsilon^2}
+\frac{1}{45}\Big(1-\frac{3r_h^2}{2l^2}\Big)
\ln\frac{\Lambda}{\epsilon}\Big] \nn \\
&&\qquad\qquad -\frac{3+(-1)^{2s}}{4}
\frac{1+2s^2}{12\pi l^2}A_h
\ln\frac{\Lambda}{\epsilon} \, ,
\eea
where $A_h = 4\pi r_h^2$.

It is not difficult to find that the entropy still depends on the square of the spins of
particles unless in the Schwarzschild and Reissner-Nordstr\"{o}m black hole cases. The
existence of a positive cosmological constant will decrease the entropy whereas the effect
of a negative one tends to increase it, vice versa. This result is in accordance with the
entropy of a massive scalar field for the Schwarzschild-anti de Sitter black hole by using
the original BWM in Ref. \cite{BW-AdS} where a negative cosmological constant is assumed
and if the square mass of the scalar field $\mu^2$ is replaced by $2/l^2$.

\textbf{Case II}: In the Kerr black hole case, the cosmological constant $\tilde{\Lambda}$
goes to zero (or in the $l\to \infty$ limit), the improved thin-layer model degenerates
to the original BWM by simply setting the infrared cutoff $L = N\varepsilon$ without any
change in the ultraviolet cutoff $\epsilon$. By the equality $N = \Lambda^2/\epsilon^2$
and the relation $\varepsilon \approx \eta^2 \simeq \epsilon^2$, the infrared cutoff
$L \approx N\epsilon^2 \simeq \Lambda^2$ is consistent with the definition $\Lambda^2 =
L\epsilon^2/\varepsilon$ as given in Refs. \cite{MS0,JY,ALO}. (There is a minor difference
in the definition of the infrared cutoff $L$ because $N$ can be sufficient large, here
it is a shift by $r_h$ from the original infrared cutoff introduced by 't Hooft \cite{GtH}).
It should be noted that the system in the brick wall region is, in fact, still in a local
thermal equilibrium, and the thin-layer model still works very well. Taking these into
account, the expression of the brick wall entropy in the Kerr black hole case reduces to
\bea
&&S/g_s = \frac{15+(-1)^{2s}}{16}\Big(\frac{A_h}{48\pi\epsilon^2}
+\frac{1}{45}\ln\frac{\Lambda}{\epsilon}\Big) \nn \\
&&\qquad\quad +\frac{3+(-1)^{2s}}{4}
\frac{s^2}{12}\Big[\frac{a^2-r_h^2}{r_h^2} +\frac{(r_h^2+a^2)^2}{ar_h^3}
\arctan\big(\frac{a}{r_h}\big)\Big]\ln\frac{\Lambda}{\epsilon} \, ,
\eea
where $A_h = 4\pi(r_h^2+a^2)$.

This result is very similar to that presented in Refs. \cite{JY,ALO}. In the scalar field
case, it coincides with the result given by Ref. \cite{MS0}. In the cases of $s = 1/2$ (Dirac
field), $s = 1$ (Maxwell field), $s = 3/2$ (Rarita-Schwinger field), and $s =2$ (linearized
gravitational field), it has the same coefficients $s^2/6$ as that in Ref. \cite{JY}. But
these coefficient terms have a different tendency because they differ by the terms in the square
bracket. In the Rarita-Schwinger field ($s = 3/2$) case, our result manifests that the spins
will increase the entropies as pointed out by Ref. \cite{ALO}. Again they differ by the
coefficients of the quadratic term $s^2$. The expression $P = s^2a^2\cda\big[\Delta-(r^2+a^2)
-\Sigma\big]/\Sigma^3$ used in Refs. \cite{JY,ALO} but written in our notations, is probably
the result that those authors had missed some terms in the process of their approximation.
We point out that it should read $P = 4s^2a^2\cda\big[\Delta-(r^2+a^2)\big]/\Sigma^3$, and
these coefficients presented in Refs. \cite{JY,ALO} are incorrect in the Kerr black hole case.
The difference of the coefficients of the quadratic term $s^2$ (the pure spin-dependent
term) between ours and those in Refs. \cite{JY,ALO} is compared in Fig. 2.

\begin{figure}[htbp]
\begin{center}
 \epsfig{figure=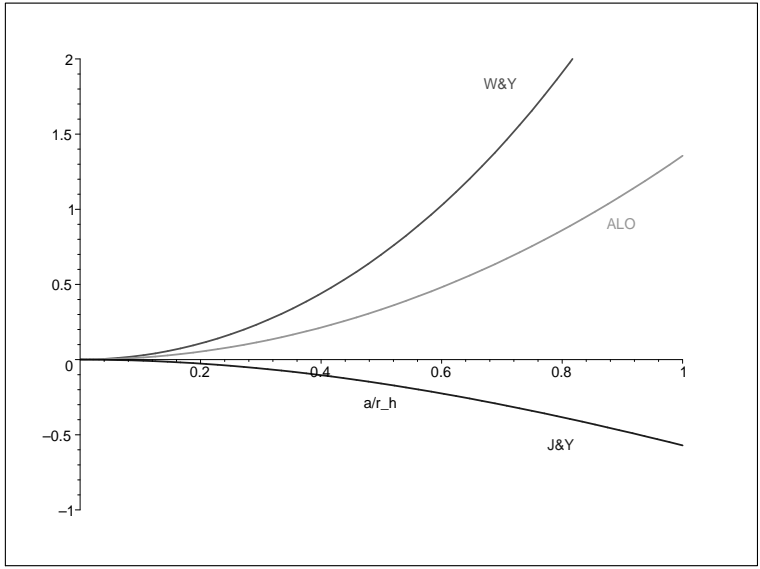,width=7.5cm,height=10cm,angle=-90}
 \caption{Compare the coefficient of the pure spin-dependent term
 to the logarithmic correction. The curve denoted by J\&Y \cite{JY}
 demonstrates that the entropy is decreased by the contribution from
 the $s^2$ term. Both the curve denoted by ALO \cite{ALO} and that of
 ours denoted by W\&Y illuminate that the spin-rotation coupling effect
 will increase the entropy of a Kerr black hole.}
\end{center}
\end{figure}

\textbf{Case III}: In the lukewarm case \cite{MM,BM}, the non-extreme Kerr-de Sitter
spacetime is characterized by the condition $M^2 = a^2\chi^2$. In the absence of rotation,
the lukewarm solution reduces to de Sitter spacetime. When the cosmological constant
vanishes, it degenerates to the extreme Kerr black hole. Obviously the lukewarm solution
is a special case of what we have discussed in the above. In this case, the cosmological
and outer black hole horizons are distinct, but the temperatures on both horizon are equal,
so the cosmological and black hole event horizons are in thermal equilibrium. Just as the
Kerr black hole case, here one need not assume local thermal equilibrium and he can still
work with the original BWM to calculate the entropy of each horizon. Of course, the cutoff
$L$ cannot be so large as to close to the cosmological horizon. The final result of the
entropy is still expressed by Eq. (\ref{eq32}) with the special restriction (namely, $M^2
= a^2\chi^2$)on the allowed range of parameter of the lukewarm solution. It should be noted
that the same result can also be arrived at by using the thin-layer method. This issue can
be easily understood as the brick wall entropy is mainly attributed to the degrees of freedom
of the field in the near horizon region. As such we think the thin-layer method is more
universal than the original BWM.

\textbf{(f)} We do not consider here the contribution from the super-radiant modes. As is well
known, for rotating black holes in an asymptotically flat space, classical super-radiance effects
occur only for bosonic but not fermionic fields \cite{SC}, however the quantum analog of
super-radiance does occur for both bosonic and fermionic fields \cite{UW}. It has been pointed
out that the super-radiant modes have a contribution to the entropy of a rotating black hole
\cite{SRE}. In the Kerr black hole case, the effect of the super-radiant modes is to halve the
leading contribution to the entropy of scalar and Dirac fields \cite{HKP,KN}. Their affect on
the subleading term to the entropy for the non-extreme Kerr-de Sitter black hole is an
interesting problem and deserves to be discussed elsewhere.

\textbf{(g)} We do not address here the renormalization of the divergence in the entropy of
the non-extreme Kerr-de Sitter black hole. A generally accepted belief is that the matter
field contributions to the entropy can be interpreted as one-loop corrections to classical
Bekenstein-Hawking entropy. The divergence of the brick wall entropy has the same origin as
the divergence of the one-loop effective action in quantum field theory in curved space. The
leading quadratic divergence of the entropy calculated by the brick wall model can be absorbed
by the renormalization of the Newton gravitational coupling constant in the one-loop effective
action for matter and gravitation fields \cite{SUBELF}. When gravity is described by a higher
curvature effective action, the standard Bekenstein-Hawking result is only the leading
contribution in the entropy, and there are still additional corrections from the higher
curvature interaction \cite{WJK}. On the other hand, quantum corrections in curved space
are known to result in higher order curvature contributions to the Einstein-Hilbert action
\cite{BD}. The subleading order logarithmic divergence in the entropy requires the introduction
in the gravitational action of term quadratic in the curvature to be renormalized \cite{FSSFF}
(see also \cite{BW-AdS,SNS,DLM,KKSYL}). The standard renormalization of Newton gravitational
coupling constant, the cosmological constant and other coefficients by the curvature squared
($R^2$) terms in the one-loop effective gravitational action \cite{BD} can remove all the
divergent (quadratic and logarithmic) terms in the entropy \cite{FSSFF}, so the remaining
quantity is finite.

In the case of non-extreme Kerr-de Sitter geometry, it is clear that the divergence coming
from the integral in the dragged optical space can be completely removed by the above procedure,
but it is unclear to us whether the logarithmic divergence arising from the effective potential
including the quadratic spin terms needs further introduction of other interaction terms to be
renormalized because this divergence explicitly contains quantum corrections to the brick wall
entropy from the spin-rotation coupling interaction. Though the renormalization of the divergence
in the brick wall entropy of the non-extreme Kerr-de Sitter black hole has not been discussed here,
we hope the expression of the entropy obtained here can shed light on this subject because there
is still little work on dealing with this problem in a rotating black hole spacetime.

\section*{V. CONCLUSIONS}

In summary, we have studied the statistical entropies of the non-extreme Kerr-de Sitter black holes
due to arbitrary spin fields, especially the subleading corrections to the black hole entropy
arising from the coupling of the spins of particles with the rotation of the black holes. First,
the null-tetrad in the Newman-Penrose formalism is introduced to decouple the Teukolsky master
equations governing massless scalar, neutrino, electromagnetic, gravitino, and gravitational field
perturbations of the Kerr-de Sitter space. Then, starting from the Teukolsky master equations we
seek the total number of the modes of the fields by taking the WKB approximation. Last, the free
energy and the quantum entropy of a non-extreme Kerr-de Sitter black hole due to arbitrary spin
fields are calculated by the improved thin-layer brick wall model. It is shown that the subleading
order contribution to the entropy is dependent on the square of three quantities, the spins of
particles, the specific angular momentum of black holes, and the cosmological constant. The
contribution of the spins of particles to the logarithmic terms of the entropy depends on the
spin-rotation coupling effect and the effect of the cosmological constant. It should be noted
that the final result also holds true in the lukewarm case where local thermal equilibrium need
not be assumed.

In particular, we have carefully investigated the effect of a positive cosmological constant
on the black hole entropy and shown that the correction from the effective potential is of the
same order as that from the integral in the dragged optical space, and both of them cannot be
discarded. However it should be noted that the correction from the ``spin potential'' $W$ has
been attributed to that of the vacuum surrounding the black holes and it has been neglected
here. Possible new quantum effects related to the ``$s W$'' term in a Kerr-de Sitter black hole
geometry may be a very interesting thing and deserve to be further investigated. It is also
needed to extend this analysis to the Kerr-Newman-de Sitter black hole case.

\section*{ACKNOWLEDGMENTS}

One of us (S.Q.W.) would like to thank Professor J. L. Jing for helpful discussions. He also thanks
Professor N. Ohta from Osaka University for his comments when this work has been done. This project
is partially supported by China Postdoctoral Science Foundation and K.C. Wong Education Foundation,
Hong Kong. M.L.Y. has been partially supported by NSF of China under Grant No. 90103002. Parts of
mathematical calculations were done when S.Q.W. stayed at the Institute of Particle Physics,
Hua-Zhong Normal University, Wuhan. Some results by manual manipulations and their confirmations
by MAPLE 7 have been done for several times while preparing this paper.

\section*{APPENDIX A. ~TEUKOLSKY-STAROBINSKY IDENTITIES}
\setcounter{equation}{0}
\renewcommand{\theequation}{A\arabic{equation}}

The Teukolsky's master equations (\ref{eq10}) and (\ref{eq11}) governing the perturbation
of Kerr-de Sitter space with massless fields can be separated into the angular parts
\bea
&&\Big[\rda\dL_{1-s}\rda\cL_s -2(2s-1)\Big(\chi\omega a\cta
+(s-1)\frac{a^2}{l^2}\cda\Big) +\lambda_s\Big]S_s = 0 \, , \nn \\
&&\Big[\rda\cL_{1-s}\rda\dL_s +2(2s-1)\Big(\chi\omega a\cta
-(s-1)\frac{a^2}{l^2}\cda\Big) +\lambda_s\Big]S_{-s} = 0 \, , \quad
\eea
and the radial parts
\bea
&&\Big[\Delta_r\cD_{1-s}\dD_0 +2(2s-1)\Big(i\chi\omega r
-(s-1)\frac{r^2}{l^2}\Big) -\lambda_s\Big](\Delta_r^sR_s) = 0 \, , \nn \\
&&\Big[\Delta_r\dD_{1-s}\cD_0 -2(2s-1)\Big(i\chi\omega r
+(s-1)\frac{r^2}{l^2}\Big) -\lambda_s\Big]R_{-s} = 0 \, .
\eea
Here we only consider the radial equations. It is clear that $(\Delta_r^sR_s)$ and $R_{-s}$
are proportional to complex conjugate functions. The exact relationship between these functions
are called the famous Teukolsky-Starobinsky identities \cite{TP,TS1,TS2}
\bea
\Delta_r^s\cD_0^{2s}R_{-s} = C_s(\Delta_r^sR_s) \, , \qquad
\Delta_r^s\cD_0^{\dagger 2s}(\Delta_r^sR_s) = C_s^*R_{-s} \, ,
\eea
with the coefficients
\bea
&& |C_{1/2}|^2 = \lambda_{1/2} \, ,  \nn \\
&&~~ |C_1|^2 = \lambda_1^2 -4\chi^2\omega^2 \alpha^2 \, ,  \nn \\
&& |C_{3/2}|^2 = (\lambda_{3/2}^2+4a^2/l^2)(\lambda_{3/2}+1-a^2/l^2) \nn \\
&&\qquad\qquad -16\chi^2\omega^2(\lambda_{3/2}\alpha^2+\alpha^4/l^2-a^2)
-4M^2/l^2 \, , \nn \\
&& |C_2|^2 = \big[\lambda_2(\lambda_2+2-2a^2/l^2)+12a^2/l^2\big]^2
-8\chi^2\omega^2\lambda_2\big[(5\lambda_2+6 -6a^2/l^2 \nn \\
&&\qquad\qquad +12\alpha^2/l^2)\alpha^2-12a^2\big]
+144\chi^2\omega^2(\chi^2\omega^2\alpha^4+2a^2\alpha^2/l^2+M^2) \, ,
\eea
where $\alpha^2 = a^2-ma/\omega$. The coefficient $|C_2|^2$ corrects the previous results
\cite{KCM}, while the appearance of $|C_{3/2}|^2$ is the first time, to our knowledge.

\section*{APPENDIX B. ~INTEGRALS IN TERMS OF THIN-LAYER BRICK WALL MODEL}
\setcounter{equation}{0}
\renewcommand{\theequation}{B\arabic{equation}}

Distinguished from the original BWM, the thin-layer BWM suggests that the entropy of a black
hole with two horizons mainly comes from a very thin layer in the vicinity of the horizon where
exists a local thermal equilibrium. Just as the original BWM, it also impose a small ultraviolet
cutoff $\varepsilon$ such that
\be
\Psi(x) = 0 \qquad \mbox{for} \quad r \leq r_h+\varepsilon \, ,
\ee
where $r_h$ denotes one coordinate of the two horizons of the non-extreme Kerr-de
Sitter black hole. In this paper, it represents the outer black hole event horizon
or the cosmological horizon, satisfying the horizon equation
\be
\Delta_{r_h} = (r_h^2+a^2)\Big(1-\frac{r_h^2}{l^2}\Big) -2Mr_h = 0 \, .
\ee
To remove the infrared divergence, it however introduces another cutoff parameter
--- an arbitrary big integer $N$ such that
\be
\Psi(x) = 0 \qquad \mbox{for} \quad r \geq r_h+N\varepsilon \, .
\ee

Suppose that the quantum field is rotating with the angular velocity $\Omega_h =
a/(r_h^2 +a^2)$ in the thin layer near the horizon of the Kerr-de Sitter black hole,
we may expand $\Delta_r$ close to the event horizon $r_h$ as
\be
\Delta_r = \Delta_{r_h}^{\prime}(r-r_h)
+\frac{1}{2}\Delta_{r_h}^{\prime\prime}(r-r_h)^2 +\cdots \, ,
\ee
and then expand three quantities $\widetilde{g}_{tt}$, $P$, and $W$ in terms of
the surface gravity $\kappa_h = \Delta_{r_h}^{\prime}/\big(2\chi(r_h^2+a^2)\big)
= 2\pi\Delta_{r_h}^{\prime}/(\chi^2A_h)$ as follows:
\bea
&& \widetilde{g}_{tt} = \frac{\Delta_{\theta}a^2\sda(r^2-r_h^2)^2
-\Delta_r\Sigma_h^2}{\chi^2(r_h^2+a^2)^2\Sigma} \nn \\
&&\quad \approx \frac{-2\kappa_h\Sigma_h(r-r_h)}{\chi(r_h^2+a^2)}
\Big[1-\Big(\frac{2r_h}{\Sigma_h}
-\frac{\Delta_{r_h}^{\prime\prime}}{2\Delta_{r_h}^{\prime}}
+\frac{4r_h^2\Delta_{\theta}a^2\sda}{\Delta_{r_h}^{\prime}\Sigma_h^2}\Big)
(r-r_h)\Big] +\cdots \, , \nn \\
&& P = \frac{4s^2+2}{l^2}+\frac{4s^2a^2\cda}{\Sigma^3}\Big[
\Delta_r-\Delta_{\theta}(r^2+a^2) +\frac{a^2\sda}{l^2}\Sigma\Big] \nn \\
&&\quad \approx \frac{4s^2+2}{l^2}+\frac{4s^2a^2\cda}{\Sigma_h^3}\Big[
\frac{a^2\sda}{l^2}\Sigma_h -\Delta_{\theta}(r_h^2+a^2)\Big] +\cdots \, , \nn \\
&& W = \frac{-a\cta}{\chi(r_h^2+a^2)\Sigma^2}\Bigg\{\Big[\chi\Sigma
+2a^2\sda\Big(1-\frac{r^2}{l^2}\Big)\Big](r^2-r_h^2) +2\Delta_r\Sigma_h\Bigg\} \nn \\
&&\quad \approx \frac{-4\kappa_ha\cta}{\Sigma_h}\Bigg\{1
+\frac{r_h}{\Delta_{r_h}^{\prime}}\Big[\chi+\frac{2a^2\sda}{\Sigma_h}\Big(1
-\frac{r_h^2}{l^2}\Big)\Big]\Bigg\}\big(r-r_h\big) +\cdots \, ,
\eea
where $\Sigma_h = r_h^2 +a^2\cda$.

Expanding the integrands in the four integrals $I_1 \sim I_4$ defined in Eq. (\ref{eq26})
\bea
&& \frac{\sqrt{-g}}{\widetilde{g}_{tt}^2}
\approx \frac{(r_h^2+a^2)^2 \sta}{4\kappa_h^2\Sigma_h}
\Big[\frac{1}{(r-r_h)^2}+\Big(\frac{6r_h}{\Sigma_h}
-\frac{\Delta_{r_h}^{\prime\prime}}{\Delta_{r_h}^{\prime}}
+\frac{8r_h^2\Delta_{\theta}a^2\sda}{\Delta_{r_h}^{\prime}
\Sigma_h^2}\Big)\frac{1}{r-r_h}\Big] +\cdots \nn \\
&&\qquad \approx \frac{(r_h^2+a^2)\sta}{4\chi\kappa_h^3\Sigma_h}
\Big[\frac{\Delta_{r_h}^{\prime}}{2(r-r_h)^2}
+\Big(\frac{3r_h\Delta_{r_h}^{\prime}}{\Sigma_h}
-\frac{1}{2}\Delta_{r_h}^{\prime\prime}
+\frac{4r_h^2\Delta_{\theta}a^2\sda}{\Sigma_h^2}\Big)
\frac{1}{r-r_h}\Big] +\cdots \, , \nn \\
&&\frac{\sqrt{-g}}{\widetilde{g}_{tt}}P
\approx \frac{-(r_h^2+a^2)\sta}{2\chi\kappa_h(r-r_h)}\Bigg\{\frac{4s^2+2}{l^2}
+4s^2a^2\cda\Big[\frac{a^2\sda}{l^2\Sigma_h^2}
-\frac{\Delta_{\theta}(r_h^2+a^2)}{\Sigma_h^3}\Big]\Bigg\}
+\cdots \, , \nn \\
&&\frac{\sqrt{-g}}{\widetilde{g}_{tt}^2}W^2
\approx \frac{4(r_h^2+a^2)^2a^2\sta\cda}{\Sigma_h^3}\Bigg\{1
+\frac{r_h}{\Delta_{r_h}^{\prime}}\Big[\chi+\frac{2a^2\sda}{\Sigma_h}\Big(1
-\frac{r_h^2}{l^2}\Big)\Big]\Bigg\}^2 +\cdots \, , \nn \\
&&\frac{\sqrt{-g}}{\widetilde{g}_{tt}}\frac{\Delta_r^{\prime\prime}}{4\Sigma}W
\approx \frac{\Delta_{r_h}^{\prime\prime}(r_h^2+a^2)a\sta\cta}{2\chi\Sigma_h^2}
\Bigg\{1+\frac{r_h}{\Delta_{r_h}^{\prime}}\Big[\chi+\frac{2a^2\sda}{\Sigma_h}
\Big(1-\frac{r_h^2}{l^2}\Big)\Big]\Bigg\} +\cdots \, ,
\eea
and carrying out the integrals with respect to $\theta$ and $r$, we finally arrive at
\bea
&&I_1 = \frac{1}{\chi\kappa_h^3}\Big[\frac{r_h^2+a^2}{\eta^2}\frac{N-1}{N}
\frac{\Sigma_h}{ar_h}\arctan\big(\frac{a}{r_h}\big)
+\Big(1-\frac{3r_h^2+a^2}{2l^2}\Big) \ln N\Big] \nn \\
&&\quad = \frac{2}{\chi\kappa_h^3}\Big[\frac{15(r_h^2+a^2)}{4\epsilon^2}
+\Big(1-\frac{3r_h^2+a^2}{2l^2}\Big) \ln\frac{\Lambda}{\epsilon}\Big] \, , \nn \\
&&I_2 = \frac{1}{4\chi\kappa_h}\Bigg\{-\frac{4+8s^2}{l^2}\Big(r_h^2+a^2\Big)
+s^2\Big[\frac{a^2-r_h^2}{r_h^2}+\frac{9r_h^2+7a^2}{l^2} \nn \\
&&\qquad +\Big(\frac{r_h^2+a^2}{r_h^2}-\frac{9r_h^2+a^2}{l^2}\Big)
\frac{r_h^2+a^2}{ar_h}\arctan\big(\frac{a}{r_h}\big)\Big]\Bigg\}\ln N \nn \\
&&~~ = \frac{1}{2\chi\kappa_h}\Bigg\{-\frac{4(r_h^2+a^2)}{l^2}
+s^2\Big[\frac{a^2-r_h^2}{r_h^2}+\frac{r_h^2-a^2}{l^2} \nn \\
&&\qquad +\Big(\frac{r_h^2+a^2}{r_h^2}-\frac{9r_h^2+a^2}{l^2}\Big)
\frac{r_h^2+a^2}{ar_h}\arctan\big(\frac{a}{r_h}\big)\Big]\Bigg\}
\ln\frac{\Lambda}{\epsilon} \, , \nn \\
&&I_3 \sim \mathcal{O}(\epsilon) \, , \qquad I_4 = 0 \, , \label{eB7}
\eea
where $A_h = 4\pi(r_h^2+a^2)/\chi$ is the horizon area. In the last step, we have replaced
the ultraviolet cutoff $\varepsilon = \eta^2 \Delta_{r_h}^{\prime}/(4\Sigma_h)$ by the proper
distance $\eta$ from the horizon to the inner brick wall $\eta = \int_{r_h}^{r_h+\varepsilon}
\sqrt{g_{rr}} dr \approx 2(\varepsilon\Sigma_h/\Delta_{r_h}^{\prime})^{1/2}$. To be comparable
with Refs. \cite{MS0,JY,ALO}, the new infrared cutoff $\Lambda$ and ultraviolet cutoff
$\epsilon$ in the above equation (\ref{eB7}) are defined by
\be
N = \Lambda^2/\epsilon^2 \, ,
\qquad \eta^2 = \frac{2\epsilon^2}{15}\frac{N-1}{N}\frac{\Sigma_h}{ar_h}
\arctan\big(\frac{a}{r_h}\big) \, .
\ee
For large $N$ and small $a$, Eq. (\ref{eB7}) implies $\eta^2 = 2\epsilon^2/15$,
which is used in the context.

\end{document}